\documentclass[]{aa}
\usepackage{graphicx}
\usepackage{txfonts}
\usepackage{natbib}
\bibpunct{(}{)}{;}{a}{}{,}

\begin{document}

\title{First radius measurements of very low mass stars with the VLTI\thanks{
Based on observations made with the European Southern Observatory
telescopes and obtained from the ESO/ST-ECF Science Archive Facility}}
 \subtitle{}
 \offprints{D. S\'egransan}

\author{D. S\'egransan \inst{1}
   \and P. Kervella  \inst{2}
   \and T. Forveille \inst{3}
   \and D. Queloz    \inst{1}
}

\offprints{D. S\'egransan \email{Damien.Segransan@obs.unige.ch}}

\institute{Observatoire de Gen\`eve, 51 chemin des maillettes, CH-1290 
           Sauverny, Switzerland
      \and European Southern Observatory, Casilla 19001, Vitacura, 
           Santiago 19, Chile
      \and Canada-France-Hawaii Telescope Corporation, P.O. Box 1597, 
           Kamuela, HI 96743, U.S.A.
}

\date{Received 15 October 2002 / Accepted 21 November 2002}

\abstract{
We present 4 very low mass stars radii measured with the VLTI using the 2.2$\mu m$ VINCI test instrument.
The observations were carried out during the commissioning of the 104-meter-baseline with two 
8-meter-telescopes. We measure angular diameters of 0.7-1.5~mas with accuracies
of 0.04-0.11~mas, and for spectral type ranging from  M0V to M5.5V.
We determine an empirical mass-radius relation for M~dwarfs based on all available radius measurements.
The observed relation agrees well with theoretical models at the present accuracy level, with possible discrepancy around 0.5-0.8~$M_{\odot}$ that needs to be confirmed.
In the near future, dozens of M 
dwarfs radii will be measured with 0.1-1\% accuracy, with the VLTI, thanks to the improvements expected from the 
near infrared instrument AMBER. 
This will bring strong observational constraints on both atmosphere and interior physics.

}

\maketitle

\section{Introduction}
Mass, radius, luminosity and chemical composition are the basic physical 
properties of a star. For a given mass and chemical
composition, theory can predict most of the stellar physical parameters 
at a given age. 
Accurate stellar mass, radius and luminosity measurements thus
provide a crucial test of our understanding of stellar physics. To be 
relevant, these parameters must be determined with an accuracy of the order of 1\% 
\citep{1991A&ARv...3...91A}.
The previously rather noisy  Mass-Luminosity relation for M~dwarfs 
has recently been greatly improved 
\citep{2000A&A...364..217D, 2003IAUS..211...}. 
By contrast, the empirical Mass-Radius relation remains poorly 
constrained, since it is based on the observations of just two of the three 
known M-type eclipsing binaries, plus three M~dwarfs radii measured 
at the Palomar Testbed Interferometer (PTI)
with accuracies of  2-4\%, \citep{2001ApJ...551L..81L}. \\
%
%
%
In this paper we present direct angular diameter measurements of four 
M dwarfs with spectral types ranging from M0V to M5.5V obtained
with the Very Large Telescope Interferometer (VLTI) and two 8-meter-telescopes 
\citep{2001Msngr.106....1G}.
The first two sections describe the observations and the data
reduction. The third section presents the visibility calibration and 
the angular diameter determination. In the last
section we compare presently known M dwarfs radiis and masses to theoretical
models.
\section{Observations}
Gl887, Gl205 and Gl191 
were observed during the first commissioning run of the VLTI with 
8-meter-telescopes 
in early November 2001, on the second and third nights after the 
``first fringes'' with these telescopes. The observations of
Gl551 
were obtained in February 2002.
\\
VINCI  is the commissioning instrument of 
the VLTI \citep{2000SPIE.4006...31K}. It uses single-mode optical fibers to recombine the light 
from two telescopes of the Paranal Observatory, and modulates the 
optical path difference around its zero position to produce 
interference fringes. This 
recombination scheme was first used in the FLUOR instrument 
\citep{1998SPIE.3350..856C} and produces high precision
visibility values, thanks to the efficient spatial filtering of 
the incoming beams and to the photometric monitoring of the filtered
wavefronts. 
\\
Transfer function calibrators (Table~\ref{tab-list_objects}) were 
selected based on their spectral type, distance to the target 
and known diameter \citep{1999AJ....117.1864C}. Calibration tests show
that a $\pm 1\,\sigma$ shift in the calibrator angular diameter values modify the 
measured M~dwarf radii by less than $0.8\,\sigma$.

The M dwarfs science targets and their calibrators
were observed in series of a few hundred interferograms, 
recorded at a frequency of 500-700 Hz. 
Due to many down times associated with the technical commissioning of
the array, the
observations of calibrator stars were not possible to the extent
usually required to
monitor extensively the transfer function. 
Nevertheless, the calibration measurements
proved sufficient to estimate the transfer function with a
good accuracy. The global interferometric efficiency of the 
VLTI mostly depends on two parameters: the atmospheric coherence time,
which varies over timescales sometimes as short as a few minutes, 
and the instrumental efficiency which changes over timescales of a 
week. Our effective calibration rate oversamples the instrumental efficiency 
variations by orders of magnitude, but risks undersampling the atmospheric
coherence time. Fortunately, the Paranal observatory site is heavily
monitored, and we could rely on continuous measurement of the atmospherical 
coherence time \citep{2000SPIE.4009..338S} to discard data intervals
when the coherence time changed significantly between the observation 
of the calibrators and science targets. \\

\begin{table}
  \center
  \tabcolsep 0.1cm
  \caption{Calibrators and instrumental+atmospherical transfer function. 
           Calibrator uniform disk angular diameters come from \citet{1999AJ....117.1864C}. 
           Internal error and external error on the transfer function are given for each calibrator. 
           Fomalhaut angular diameter was redetermined using the VLTI. 
           (*) calibrator was not used due low visibilities or poor accuracy on their diameters}
  \begin{tabular}{|cccc|lll|}
\hline\hline
          &            & Spec.&{\sc Diameter}     &\multicolumn{3}{c|}{{\sc Transfer function}}\\
Targets   & Calib.     & Type &$\theta_{UD}$[mas] & $\left|T_{a+i}\right|^{2}$& $\sigma_{tot}$& $(\sigma_{stat}/\sigma_{cal})$       \\
\hline           
GJ205&39-Eri       &K3III &1.81$\pm$0.02&0.478&0.005&(.002/.005)\\
     &HD36167$^{*}$&K5III &3.55$\pm$0.06&0.486&0.037&(.006/.036)\\

GJ887&HR8685       &M0III &2.01$\pm$0.02&0.448&0.005&(.003/.004)\\ 
     &Fomalhaut$^{*}$&A3V &2.13$\pm$0.06&0.433&0.013&(.004/.013)\\  

GJ191&39-Eri         &K3III &1.81$\pm$0.02&0.420&0.005&(.004/.004)\\
     &Gam02 Vol$^{*}$&K0III &2.45$\pm$0.06&0.411&0.014&(.006/.013)\\
GJ551&HD110458     &K0III &1.66$\pm$0.02&0.509&0.006&(.004/.004)\\
\hline\hline
\end{tabular}
\label{tab-list_objects}
\end{table}

\section{From fringes to visibilities}
\begin{figure}[t]
   \centering
\includegraphics[width=8cm]{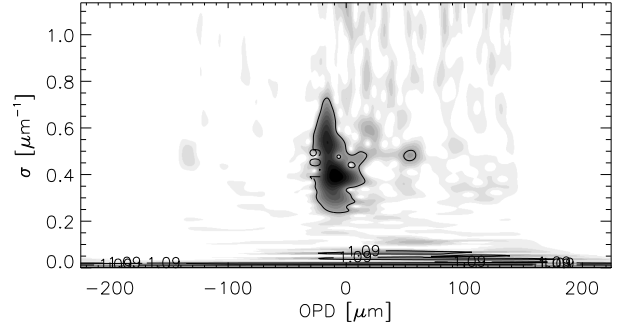}
   \caption{    Wavelet Spectral density of a pistonned interferogram. 
    The effect of strong differential piston is to distort the interferogram 
    fringe peak in the time-frequency space. Part of the interferogram's 
    energy is spread out in both time and frequency, preventing a good 
    measurement of the total energy.}
   \label{Fig-wavelets}
    \end{figure}

We used a customised version of the VLTI/VINCI data reduction pipeline,
whose general principle is based on the original FLUOR algorithm 
\citep{1997A&AS..121..379C}. Despite the high modulation frequency of 
the fringes, many recorded interferograms present a differential piston 
signature, as well as strong photometric fluctuations. The latter are
expected for large apertures without adaptive optics correction,
as individual speckle come in and out of the fiber input 
aperture. To overcome the noise amplification caused by low photometry
data points, which could strongly bias the resulting visibilities,
interferograms were calibrated by the average value of the photometry
over the fringe packet. The two calibrated output interferograms were then  
subtracted to remove residual photometric fluctuations (Kervella, in prep).\\
In parallel to the FLUOR algorithm, based on Fourier analysis, we   
implemented a time-frequency analysis \citep{1999wfoi.conf..290S} based on the 
continuous wavelet transform \citep{1992AnRFM..24..395F}. 
Instead of the projection onto a sine wave of the Fourier transform,
the wavelet transform decomposes it onto a function, {\it ie.} the Wavelet,  
that is localised in both time and frequency. We use as a basis the Morlet 
Wavelet, a gaussian enveloppe multiplied by sine wave. With the proper
choice of the number of oscillations inside the gaussian envelope,
the Morlet wavelet closely matches a VINCI interferogram. It is 
therefore extremely efficient at localizing the signal in both time 
and frequency.\\
As illustrated on figure \ref{Fig-wavelets}, differential piston strongly 
affects the amplitude and the shape of the fringe peak in the wavelet 
power spectrum. We therefore select on the shape of fringe peak in the 
time-frequency domain to efficiently reject ``pistonned'' interferograms.  
We then derive visibilities from the wavelet spectral density,
after removing the residual photon and detector noise.

\section{Angular Diameter \& Limb Darkening \label{sec-adld}}
Stellar angular diameters are usually quoted for both a uniform disk 
and a limb-darkened disk, though only the latter is directly
comparable to actual stellar models. The monochromatic visibility of 
a limb darkened disk is \citep{2000MNRAS.318..387D} :
\begin{equation}
V_{\lambda} = \frac{\int_{0}^{1}{ d\mu I\left(\mu \right) \mu J_{0}\left(\pi B \theta_{LD}/\lambda \left(1-\mu^{2}\right)^{1/2}\right)  }}{\int_{0}^{1}{ d\mu I\left(\mu \right)}}
\label{eq-vld}
\end{equation}
where $I\left(\mu \right)$ is the surface brightness, $J_{0}$ is the zero$^{th}$ order Bessel function,  $B$ the 
projected baseline, $\lambda$ the wavelength, and 
$\theta_{LD}$ the limb darkened diameter (in radian).

The VINCI instrument has no spectral resolution, and a bandpass defined
by a $K^{'}$ filter (2-2.3$\mu m$).
For marginally resolved disks ($B{\theta}/{\lambda}<<1$), Eq.~\ref{eq-vld}
remains approximately valid for polychromatic measurements, with $\lambda$
then being the effective wavelength of the system. It differs from the 
mean wavelength of the filter through the wavelength-dependency
of fiber coupling, combiner transmission, and the stellar spectrum. 
Since interferograms are Fourier transforms of the instrumental
bandpass, they provide, in principle, an accurate measurement of the effective
wavelength, obtained as the average barycenter of the fringe peak in the
spectral density space. 
We computed such effective wavelengths for each star and compared 
them to the one given by the VINCI instrument modelisation 
($\lambda_{eff}=2.195 \pm 0.010 \mu m$). Our estimations of $\lambda_{eff}$
present a large scatter which is due to atmosperical differential piston and photometric 
fluctuations. We thus used the VLTI/VINCI effective wavelength provided by ESO and 
the effective temperature of each star to compute an accurate $\lambda_{eff}$ for each 
target and calibrator.

Due to their relatively short baselines, and limited uv-coverage, these 
observations cannot discriminate between an unphysical uniform
star and a more realistic slightly larger limb-darkened star. To correct 
for limb-darkening, we therefore have to use a theoretical limb-darkening law.
We adopt the non-linear limb-darkening coefficients from 
\citet{2000A&A...363.1081C}, Eq.~\ref{eq-ld1}, based on atmospheric 
models from \citet{1997ARA&A..35..137A}. We use gravities and 
effective temperatures from  \citet{1998A&A...337..403B} isochrones at 5~Gyr 
(Table \ref{table-ldc}) to select the apropriate entry
in the \citet{2000A&A...363.1081C} tables. 
\begin{equation}
I\left(\mu \right)=I\left(1\right)\left[1-\sum_{k=1}^{4}{a_{k}(1-\mu^{k/2})}\right]
\label{eq-ld1}
\end{equation}
In Equation \ref{eq-ld1}, the $a_{k}$ are the limb-darkening coefficients and 
$\mu=cos(\gamma)$, with $\gamma$ the angle between the line of sight and the 
emergent ray.
%
%
%
   \begin{figure}
    \includegraphics[width=8cm]{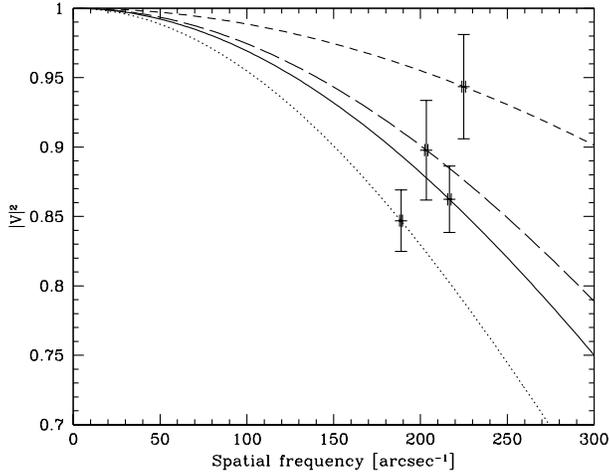}
   \caption{Measured visibilities (data points) and best-fit model (curves, 
            equation \ref{eq-ld1}) 
            for GJ 205 (plain), GJ887 (dots), Gl191 (short dash), and Gl191 
            (long dash)  
            }
   \label{Fig-V2}
    \end{figure}
The limb-darkening correction that we apply is in principle model-dependent.
Fortunately, it is also quite small for M dwarfs, only 1-2\% 
(Table~\ref{table-ldc}). The correction itself is at best marginally 
relevant at our current accuracy level for individual diameters
, and its theoretical uncertainties can for now be
safely neglected. Future observations with longer baselines, at shorter 
wavelengths, and with a more accurate calibration, will need to consider 
the issue more carefully, but they will also directly measure the 
limb-darkening law for the closest M dwarfs.\\ 

\begin{table*}
  \center
  \tabcolsep 0.1cm
  \caption{Measured angular diameter for the target stars. 
The uniform disk diameters for GJ~699, GJ~15A, GJ~411, GJ~380, GJ~105A are
from \citet{2001ApJ...551L..81L}. For consistency with our measurements, 
we recomputed their limb darkened diameters, masses and radii. The masses 
for Gl 380 and Gl105A are based on the \citet{1998A&A...337..403B} M-L
relation, with an arbitrary 5\% error bar. }
  \begin{tabular}{|r|c|cc|cc|cccc|ccc|cc|cc|cc|}
\hline
          &Spectral&    \multicolumn{2}{|c|}{{\sc Photometry \& Masses}}& \multicolumn{6}{c|}{{\sc Limb Darkening,  K band} (models)} &\multicolumn{3}{|c|}{ {\sc Diameter} [mas]}& \multicolumn{2}{|c|}{{\sc Radius} [$R_{\odot}$]}&\multicolumn{4}{|c|}{{\sc Atm. prop.}} \\
\cline{3-19}
Object    & Type   &$M_{K}$&$M/M_{\odot}$&g&$T_{eff}$&$a_{1}$&$a_{2}$&$a_{3}$&$a_{4}$ &$\theta_{UD}$  &$\theta_{LD}$ & $\sigma_{\theta}$&$R$ &$\sigma_{R}$&$T_{eff}$ &$\sigma_{T_{eff}}$&g &$\sigma_{g}$\\  
\hline
GJ205&M1.5V&5.09&0.631$\pm$0.031&4.70 & 3894&1.11&-1.11&0.92&-0.31&1.124&1.149&0.11&0.702&0.063&3520&170&4.54&0.06\\
GJ887&M0.5V&5.79&0.503$\pm$0.025&4.80 & 3645&1.61&-2.35&2.00&-0.68&1.366&1.388&0.04&0.491&0.014&3626& 56&4.76&0.03\\
GJ191&M1V  &7.08&0.281$\pm$0.014&4.98 & 3419&1.76&-2.72&2.39&-0.82&0.681&0.692&0.06&0.291&0.025&3570&156&4.96&0.13\\
GJ551&M5.5V&8.80&0.123$\pm$0.006&5.19 & 3006&1.94&-2.80&2.39&-0.81&1.023&1.044&0.08&0.145&0.011&3042&117&5.20&0.23\\

\hline\hline
GJ699&M4V  &8.21&0.158$\pm$.008         &5.11 & 3193&  1.87&  -2.88&  2.54&  -0.88&0.987&1.004&0.04&0.196 &0.008&3163&65&5.05&0.09\\
GJ15A&M2V  &6.27&0.414$\pm$.021         &4.87 & 3541&  1.66&  -2.48&  2.14&  -0.73&0.984&1.000&0.05&0.383 &0.02 &3698&95&4.89&0.06\\
GJ411&M1.5V&6.33&0.403$\pm$.020         &4.88 & 3533&  1.67&  -2.50&  2.16&  -0.73&1.413&1.436&0.03&0.393 &0.008&3570&42&4.85&0.03\\
GJ380&K7V  &4.77 &0.670$\pm$.033$^{\ast}$&4.65& 4106&  1.09&  -1.01&  0.83&  -0.28&1.268&1.155&0.04&0.605 &0.02 &-&-&4.70&0.03\\
GJ105A&K3V &4.17&0.790$\pm$.039$^{\ast}$&4.56 & 4603&  0.86&  -0.53&  0.38&  -0.13&0.914&0.936&0.07&0.708 &0.05 &-&-&4.63&0.05\\
\hline
\end{tabular}
\label{table-ldc}
\end{table*}

\section{Mass-Radius relation}
An accurate empirical mass-radius relation is an essential constraint
on stellar interior structure, evolutionnary models and atmospheric physics.
The interior structure is largely determined by the equation of state, whose
derivation for very low mass stars, brown dwarfs, and planets involves 
the complex physics of strongly correlated and partially degenerated 
quantum plasma \citep{2000ARA&A..38..337C}.\\
To obtain the empirical mass-radius relation we need to convert angular 
diameter into linear radius, which is easily done thanks to accurate HIPPARCOS
parallaxes for all sources, and we need accurate masses. For these single
stars masses can only be estimated, from IR photometric measurements   
and accurate mass-luminosity relations. Fortunately, the K-band
mass-luminosity relation has very little intrinsic dispersion for M dwarfs
. We estimated the masses 
of the stars listed in Table \ref{table-ldc} using
an update  \citep{2003IAUS..211...}  of the \citep{2000A&A...364..217D} empirical 
K band mass-luminosity relation, now based on 27 accurate masses and 
luminosities. The data dispersion around the average empirical relation 
corresponds to a mass error of $\sim$ 5\%.
Masses and radii are summarized in Table 
\ref{table-ldc}.

\begin{figure}
   \centering
    \includegraphics[width=8cm]{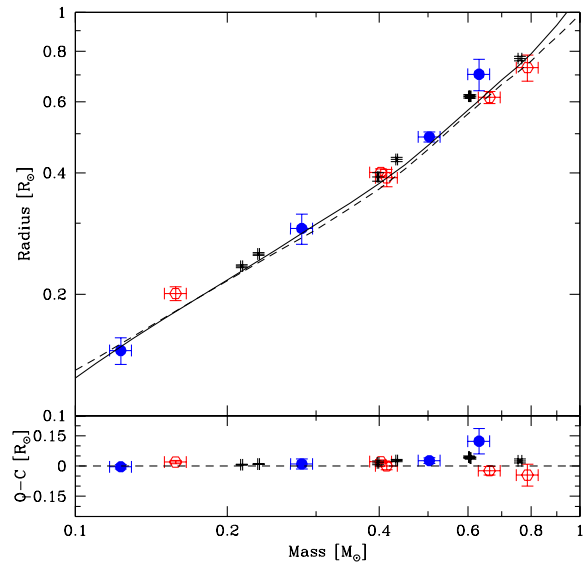}
   \caption{Comparison between observational radii \& masses measurements
     and the theoretical mass-radius relation. The
     solid and dashed curves are the 5~Gyr and 
     0.4-1~Gyr theoretical isochrones of \citet{1998A&A...337..403B}, which do
     not differ much over the present mass range. The filled
     circles are radius measurements from this paper, the open circles are 
     PTI measurements by \citet{2001ApJ...551L..81L}, and the dots are 
     masses and radii of three eclipsing
     binaries \citep{1996ApJ...456..356M, 2002ApJ...567.1140T, 2002astro.ph.11086R}. 
     The error bars and the residuals from the model are shown at the bottom 
     of the figure.}
   \label{Fig-MRR}
    \end{figure}

Figure~\ref{Fig-MRR} compares the empirical radii \& masses with 
5~Gyr and 1~Gyr theoretical isochrones from \citet{1998A&A...337..403B}.
The \citet{1998A&A...337..403B} models reproduce the observation fairly 
well in the sampled range, between 0.65 and 0.12 $M_{\odot}$.
At a more detailed level though, one notices that the models underestimate 
the radii for YY~Gem, V818~Tau and GJ~205, with masses in the 
0.5-0.65$M_{\odot}$ range.
The discrepancy is highly significant for the eclipsing 
binaries, as extensively discussed by \citet{2002ApJ...567.1140T} and is more 
marginal for long baseline interferometry data. There is an indication
in Fig.~\ref{Fig-MRR} that the model reproduces the observations well
below 0.5~$M_{\odot}$, and only become discrepant above that value. If real,
this suggests that the shortcomings of current models have to be searched
in the energy transport (convection description, opacities), rather than in 
the equation of state (EOS). As their mass decreases, stars at the bottom of 
the main sequence have increasingly simple transport properties (they are 
fully convective below $\sim$0.3~$M_{\odot}$), and an increasingly correlated 
and degenerate EOS. EOS shortcomings are therefore expected to show up most 
prominently at the lowest masses, and transport problems at higher masses.
That result obviously needs confirmation from additional data points and 
from more accurate measurements, but the consistency between the eclipsing 
binary and the resolved single stars is comforting, as these two datasets were
obtained through completely independent methods.\\
In addition to providing physical radii, the angular radii can be combined
with integrated flux measurement to obtain the effective temperature from
the Stefan-Boltzmann law. Up to now that process has typically been inversed, 
to derive radii for stars whose effective temperatures were obtained from
comparison with model spectra (e.g. \citet{2000ApJ...535..965L}). Amongst 
the sources in Table~\ref{table-ldc},
only GJ~699 has an integrated flux measurement \citep{2000ApJ...535..965L},
and we thus have to rely on the bolometric correction polynomial fit of 
\citep{2000ApJ...535..965L} to estimate it for the other targets. 

In spite of these increased uncertainties on the bolometric flux, the 
resulting effective temperatures have fairly small error bars (Table \ref{table-ldc}). 
With increasingly precise direct angular radii, and with integrated
flux measurements for all targets, this test can become a very 
discriminating validation of model atmospheres, with potential accuracies 
at the $\sim$ 1\% level on both T$_{eff}$ and log(g). 
The present diameter measurements, which have similar precisions to those
obtained at PTI \citep{2001ApJ...551L..81L}, have been obtained during 
the commissioning phase of the VLTI interferometer, with a recombiner that
was specified to validate and debug the interferometer, and not for an optimum
scientific output. With the improvements expected from the near infrared scientific
recombiner, AMBER \citep{2001sf2a.conf..615P}, from adaptive optics on the unit telescopes,
the VLTI will measure dozens of M dwarfs with 0.1-1 \% accuracy. This will 
bring very strong observational constraints on models of both stellar interiors
and atmospheres. 

\begin{acknowledgements}
This research  has made use 
of the SIMBAD  database operated by CDS, Strasbourg, France.
D. S.  acknowledges the support of the \emph{Fonds National Suisse de la 
Recherche Scientifique}.
\end{acknowledgements}

\bibliographystyle{aa}
\bibliography{Ej152.bib}
\end{document}